# Large spin-Hall effect in Si at room temperature

Paul C. Lou[1], Anand Katailiha[1], Ravindra G Bhardwaj[1], Tonmoy Bhowmick[2], W.P. Beyermann[3], Roger K. Lake[2,4] and Sandeep Kumar[1,4*]

[1] Department of Mechanical Engineering, University of California, Riverside, CA 92521, USA

[2] Department of Electrical Engineering and Computer Science, University of California, Riverside, CA 92521, USA

[3] Department of Physics and Astronomy, University of California, Riverside, CA 92521, USA

[4] Materials Science and Engineering Program, University of California, Riverside, CA 92521, USA

**Abstract**

Silicon's weak intrinsic spin-orbit coupling and centrosymmetric crystal structure are a critical bottleneck to the development of Si spintronics, because they lead to an insignificant spin-Hall effect (spin current generation) and inverse spin-Hall effect (spin current detection). Here, we undertake current, magnetic field, crystallography dependent magnetoresistance and magneto-thermal transport measurements to study the spin transport behavior in freestanding Si thin films. We observe a large spin-Hall magnetoresistance in both p-Si and n-Si at room temperature and it is an order of magnitude larger than that of Pt. One explanation of the unexpectedly large and efficient spin-Hall effect is spin-phonon coupling instead of spin-orbit coupling. The macroscopic origin of the spin-phonon coupling can be large strain gradients that can exist in the freestanding Si films. This discovery in a light, earth abundant and centrosymmetric material opens a new path of strain engineering to achieve spin dependent properties in technologically highly-developed materials.

I.   Introduction

The spin-Hall effect (SHE)[1,2] and its reciprocal is an efficient mechanism of generation and detection of spin current, which arises in materials with large intrinsic spin-orbit coupling (SOC). However, large SOC can also arise due to broken inversion symmetry. In case of centrosymmetric materials, the symmetry can be altered using inhomogeneous strain [2-6]. The broken symmetry in centrosymmetric materials will give rise to a flexoelectric polarization due to an inhomogeneous strain field as shown in Figure 1 (a)[7]. Recently, the flexoelectric effect[8-11] due to a strain gradient has been experimentally observed in centrosymmetric Si[12], which provides a foundation for this study. Based on the flexoelectric coefficient reported for Si[13], the strain and strain gradient mediated Rashba Dresselhaus SOC may lead to SHE in Si with a magnitude similar to that of GaAs (Supplementary information-A[14] and see, also, references [3,9,13,15-20] therein). Traditionally, strain gradient experiments involve bending thin films on soft substrates[12]. Alternatively, a freestanding beam will buckle automatically due to residual stresses. The stresses and, as a consequence, the buckling can be controlled using thermal expansion. Within this framework, we perform experimental measurements of SHE in Si (p-doped and n-doped) free-standing thin films. Using spin-Hall magnetoresistance (SMR), magnetoresistance (MR) as a function of crystallographic direction and magneto-thermal transport measurements, we report an unexpectedly large SHE that is comparable to or larger than those found in Pt.

II.  Experimental setup

Using standard micro/nanofabrication techniques (Supplementary information-B[14]), we fabricated a freestanding, multilayer thin film structure with a four-probe longitudinal resistance setup as shown in Figures 1 (b,c). The false color scanning electron micrograph in Figure 1 (c) shows the fabricated experimental device geometry[21,22]. The length and width of the suspended beam are 160 μm and 12 μm, respectively. The materials and thicknesses of the multilayer thin film are Pd (1 nm)/Ni$_{80}$Fe$_{20}$(25 nm)/MgO (1.8 nm)/p-Si (2 μm).

There are two contributions to the strain and strain gradient in a freestanding thin film, residual thermal expansion strain due to the thin film processing and buckling strain due to the removal of the substrate. The strain profile in the specimen will be superposition of a uniform normal strain due to thermal expansion and a strain gradient due to buckling as shown in Figure 1 (b). Residual stresses in thin films on a substrate may cause strain gradients, but they are not controlled. However, the buckling of a freestanding thin film will change with an increase in applied current mediated by Joule heating. If strain is a primary driving mechanism, the spin transport behavior will change as a function of current. In an all-metal system, the OP-AMR and SMR were shown to be functions of temperature [23] with a cross-over occurring below room temperature. Thus, Joule heating can result in pure temperature effects combined with temperature driven strain effects.

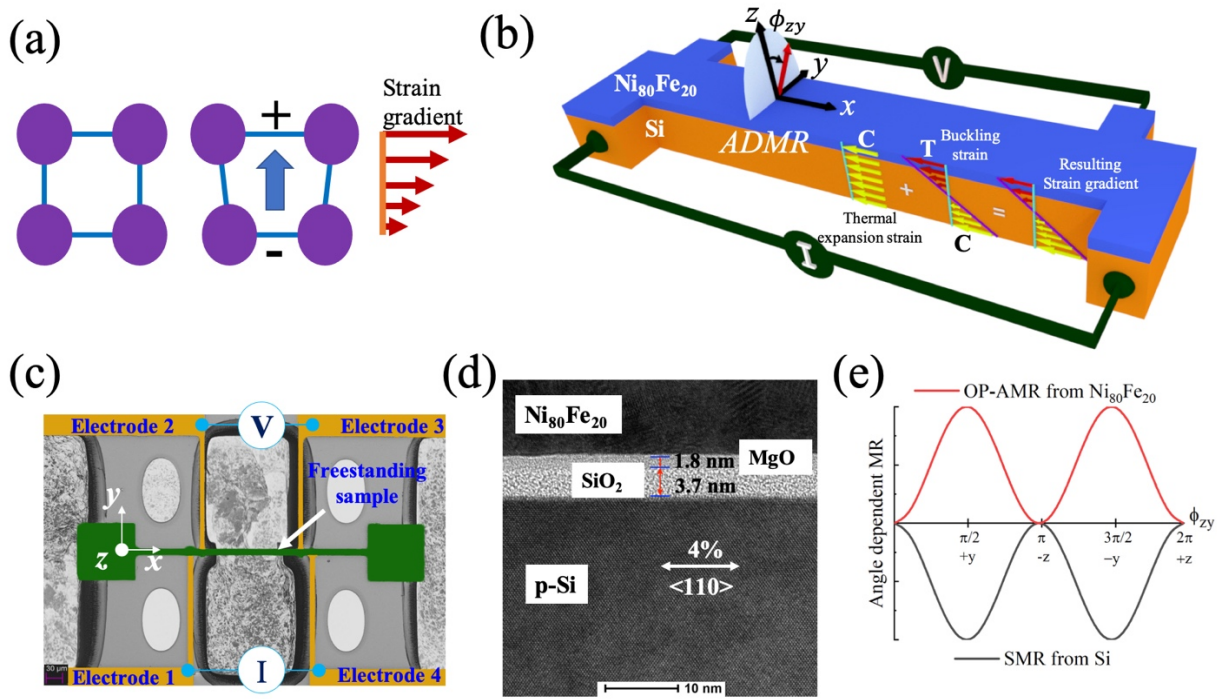

Figure 1. Illustration of the test device, the effect of strain, and competing magnetoresistance effects. (a) Flexoelectric polarization due to a strain gradient. (b) Device schematic and experimental setup for angle dependent magnetoresistance measurements in the yz-plane. The strain gradient due to thermal expansion and buckling is also shown. C denotes compression and T denotes tension. (c) A false color SEM micrograph showing the representative experimental device with freestanding channel, (d) high resolution transmission electron micrograph showing the thin film structure at the Si interface and the estimated strain in <110> direction. (e) Expected symmetry behavior of the magnetoresistance corresponding to OP-AMR and SMR.

To estimate the strain in the Si near the interface, we made a similar device with a longer 600 μm Si beam so that we could measure the buckling deformation (Supplementary Figure S1) and estimate the residual stresses (Supplementary information–C[14]). From high resolution transmission electron microscope (HRTEM) diffraction along <110> and <111> directions, we estimate 4% tensile strain near the interface as shown in Figure 1 (d) and Supplementary Figure S2. This calculated stress is less than the fracture stress of single crystal Si[24]. While the strain is estimated for very small region of the thin film, but symmetry of beam bending helps us in estimating the strain gradient. It is noted that the HRTEM sample preparation may release some of the stresses leading to underestimation of strain. The HRTEM image also shows the presence of a native oxide (~3.7 nm) in spite of Ar milling. However, the oxygen deficient native oxide will have dangling bonds and pin-holes that allow spin dependent electron tunneling and indirect exchange interactions required for spin transport studies. HRTEM imaging is complemented by energy-dispersive X-ray spectroscopy (EDS) elemental mapping, which shows absence of any Ni or Fe diffusion in Si layer as shown in Supplementary Figure S3 (Supplementary information – D[14]).

## III. Results
### A. SMR measurement in p-Si thin film

SMR is a widely used characterization technique to identify SHE[25-27]. For SMR characterization, an angle dependent magnetoresistance (ADMR) measurement is carried out on a bilayer specimen that consists of a ferromagnetic (FM) layer and a normal layer. In our case, the normal layer is the Si layer. If SHE exists in the Si layer, the spin current absorption and reflection at the FM interface depends on the angle of the externally applied magnetic field. The spin absorption and reflection at the FM interface then modulates the longitudinal resistance of the Si through the inverse SHE. The SMR behavior is identified by field rotation in the yz-plane (field perpendicular to the direction of current) as shown in Figure 1 (b).

The $Ni_{80}Fe_{20}$ thin film exhibits an out of plane anisotropic magnetoresistance (OP-AMR)[28] in the yz-plane due to dimensional confinement. Hence, the total magnetoresistance (MR) of the multilayer film will be a superposition of SMR from the Si layer and OP-AMR from the $Ni_{80}Fe_{20}$ as shown in Figure 1 (e). The angular resistance modulation (in the yz-plane) due to OP-AMR and SMR can be written as,

$$R = R_0 + (\Delta R_{OP-AMR} - \Delta R_{SMR})m_y^2 , \qquad (1)$$

where $R_0$ is the base resistance, $\Delta R_{OP-AMR}$ is the modulation in resistance due to OP-AMR, $\Delta R_{SMR}$ is modulation in resistance due to SMR, and $m_y$ is the magnetic moment projection along the y-axis defined in Fig. 1 (b,c). Using ADMR measurements in the yz-plane, the SHE behavior can be distinguished from the OP-AMR contribution due to their opposite symmetries as shown in Figure 1 (e).

The ADMR measurements are performed at a constant magnetic field of 4 T and as a function of applied current from 100 µA to 2 mA as shown in Figure 2 (a). The p-Si layer in the suspended structure is oriented along the <110> direction. At 100 µA, we observe an ADMR behavior having a polarity similar to OP-AMR as shown in Figure 1 (e) and 2 (a). At 500 µA, the MR behavior is minimal and a further increase in current to 2 mA leads to a change in polarity that can be attributed to the contribution from SMR dominating the OP-AMR. Due to competition between OP-AMR and SMR, the measured ADMR values are small and are reaching the limit of instrumental resolution as observed in Figure 2 (a). However, the results presented in Figure 2 (a) are not artifact due to instrumental resolution since they are supported by further measurements presented in this study.

To ensure that observed behavior is due to interlayer spin dependent interactions, we measured the MR of the specimen as a function of magnetic field applied along y-axis and z-axis as shown in Supplementary Figure S4 and magnetic hysteresis measurement as shown in Supplementary Figure S5 (Supplementary information-E[14] and see, also, references [29-31]). The measurement clearly shows a spin valve behavior due to spin dependent interactions across the layers in spite of thick oxide layer.

To demonstrate the competition between SMR and OP-AMR, the ADMR measurement is carried out by keeping the current constant at 900 µA while increasing the magnetic field from 1 T to 10 T, as shown in 2 (b). At low fields, the ADMR behavior displays polarity similar to SMR, indicating that magnitude of SMR is larger than OP-AMR. The ADMR is minimal at 6 T and changes polarity with further increase in strength of applied magnetic field to 10 T. The OP-AMR is a function of magnetic field due to magnon MR while SMR is not. The magnetic field dependence of the ADMR is consistent with the picture of the two competing mechanisms of OP-AMR and SMR. To further support our argument, we measured the MR as a function of magnetic field from 14 T to -14 T as shown in Supplementary Figure S6 (Supplementary information-E[14] and see, also, references [29-31]). The measurement shows a transition from SMR to OP-AMR behavior around 6 T, which supports ADMR measurement. This behavior arises due to diverging slopes of high field magnon MR [31] for field applied along y-axis and z-axis. Since the slope of magnon MR in z-axis is larger, it supports our assertion that interlayer spin dependent interactions are responsible for SMR symmetry observed in ADMR measurements.

The observed SMR behavior can be quantified using thickness dependent measurements. However, unlike a deposited thin film, a single crystal Si layer makes thickness dependent measurements difficult. For quantitative estimation of the SMR, we calculate the maximum amplitude of the ADMR at each current using a sine square curve fit. The $Ni_{80}Fe_{20}$ resistance ($\rho_{Ni_{80}Fe_{20}} = 5.43 \times 10^{-7}$ Ωm) is measured from a control specimen and the resistance ($\rho_{p-Si} = 5.25 \times 10^{-5}$ Ωm) of the p-Si layer is estimated using a parallel resistor model. This value is consistent with the bulk SOI wafer resistivity of 0.001 – 0.005 Ω cm. With these resistivity values, 56% of the current flows in the $Ni_{80}Fe_{20}$ layer and 44% flows in the p-Si layer. Then, we measured ADMR in a $Ni_{80}Fe_{20}$ control sample as shown in Supplementary Figure S7 to evaluate the OP-AMR contribution. The OP-AMR measurement in the $Ni_{80}Fe_{20}$ control specimen clearly shows that non-linear effects due to high field are not the cause of SMR like symmetry behavior in p-Si sample. We fabricated a second control sample with 25 nm of $SiO_2$ in between the $Ni_{80}Fe_{20}$ and p-Si layers. The ADMR measurement on this sample also displays OP-AMR response as shown in Supplementary Figure S8. To further support our work, we fabricated a third control sample with $Ni_{80}Fe_{20}$ (25 nm) on a freestanding oxide membrane. This sample also exhibits OP-AMR response as shown in Supplementary Figure S9. These control experiments clearly show that the observed SMR behavior arises due to the p-Si layer (Supplementary Section F[14]). From OP-AMR measurements on a $Ni_{80}Fe_{20}$ control specimen, we estimate the OP-AMR to be 0.125% at 4 T for the multilayer structure. Using this value for the OP-AMR, the magnitude of the SMR is 1.15 x $10^{-3}$ at 1.25 mA. It is an order of magnitude larger than that of Pt[25], and it is of same order as the SMR reported in some topological insulators[32,33]. To approximate the spin-Hall angle ($\Theta_{SH}$), we utilize the SMR equations for a bimetallic[26] structure (the full expression is given in Supplementary information-H[14] and see, also, reference [34]). For our geometry and materials, it simplifies to

$$\frac{\Delta R_{xx}^{SMR}}{R_{xx}^0} \approx -\Theta_{SH}^2 \frac{\lambda_N}{d} \frac{2*tanh^2\left(\frac{d}{2\lambda_N}\right)}{(1+\xi)coth\left(\frac{d}{\lambda_N}\right)} \approx -\Theta_{SH}^2 \frac{2\lambda_N}{(1+\xi)d} \quad . \tag{2}$$

For the measured values of $\frac{\Delta R_{xx}^{SMR}}{R_{xx}^0} = 7.88 \times 10^{-4} - 0.00115$, a p-Si spin diffusion length of $\lambda_N = 310$ nm[35], a Si layer thickness of $d = 2$ μm, and a current shunting parameter $\xi = 1.21$, the spin Hall angle is $\Theta_{SH} = 0.075$-$0.096$. This value is three orders of magnitude larger than $\Theta_{SH}=10^{-4}$ reported previously for p-Si[36] and it is of same order as the spin-Hall angle of Pt ($\Theta_{SH} = 0.055 - 0.1$) [25,37]. 310 nm is one of the largest values of spin diffusion length reported for p-Si, and it results in a lower bound on the spin-Hall angle. Shikoh et al. [38] reported a spin diffusion length of 148 nm in p-Si whereas Weng et al. [39] reported a value of 40 nm. Using these values increases the estimated spin-Hall angles to 0.26. Hence, the extracted values of the spin-Hall angle for p-Si in our samples can be in the range of 0.075-0.26. These values are two orders of magnitude larger than GaAs are not expected to arise due to strain gradient only (Supplementary information-A[14]). However, the strain may be the underlying cause of the large variance in reported spin-Hall angle values.

### B. Piezoresistive effects in p-Si

Similar to SMR, the longitudinal resistance of the specimen ($R_{xx}^0$), estimated from ADMR measurements, is also a function of applied current, as shown in Figure 2 (c). The resistance decreases as the applied electric current is increased from 100 μA to 1.25 mA, and then it increases after 1.25 mA. The Joule heating due to the current leads to an increase in the sample temperature, which consequently causes thermal expansion induced stresses. P-Si[40], $Ni_{80}Fe_{20}$, and the composite multilayer[30] have positive temperature coefficients of resistance (TCR). We measured the resistance of the specimen as a function of chamber temperature from 300 K to 350 K as shown in Supplementary Figure S10, which clearly shows an increase in resistance as a function of temperature. The rise in chamber temperature does not increase the buckling stresses significantly since the substrate is also expanding. In contrast, an increase in current causes thermal expansion of the sample structure only since the substrate (heat sink) temperature is not changing. Hence, the resistance of the multilayer thin film should not decrease due to Joule heating. However, thermal expansion induced compressive stresses along the Si <110> direction leads to a decrease in the resistance attributed to piezoresistance[41-43]. For the freestanding structure in this study, the decrease in resistance from piezoresistance is larger than the increase in resistance due to the temperature rise for applied currents less than 1.25 mA. From the parallel resistor model, we estimate that the electrical resistance of the p-Si layer changes from ~360.5 Ω to ~349.3 Ω. The bulk piezoresistance coefficient for p-Si <110> is 71.8×10$^{-11}$ m$^2$/N[42,43] and it changes by a factor of ~0.4[42] for doping concentrations above 10$^{19}$ cm$^{-3}$. Using the piezoresistance coefficients, we estimate the change in compressive stress due to 1.25 mA of heating current is ~107 MPa in addition to residual stresses prior to Joule heating. The change in compressive stress of 107 MPa is larger than the estimated buckling stress (~20 MPa) of the specimen, which will enhance the existing strain gradient due to residual stresses. This analysis shows that strain and strain gradient is the underlying reason for increase in the magnitude of SMR observed in the current dependent ADMR measurements shown in Figure 2 (a).

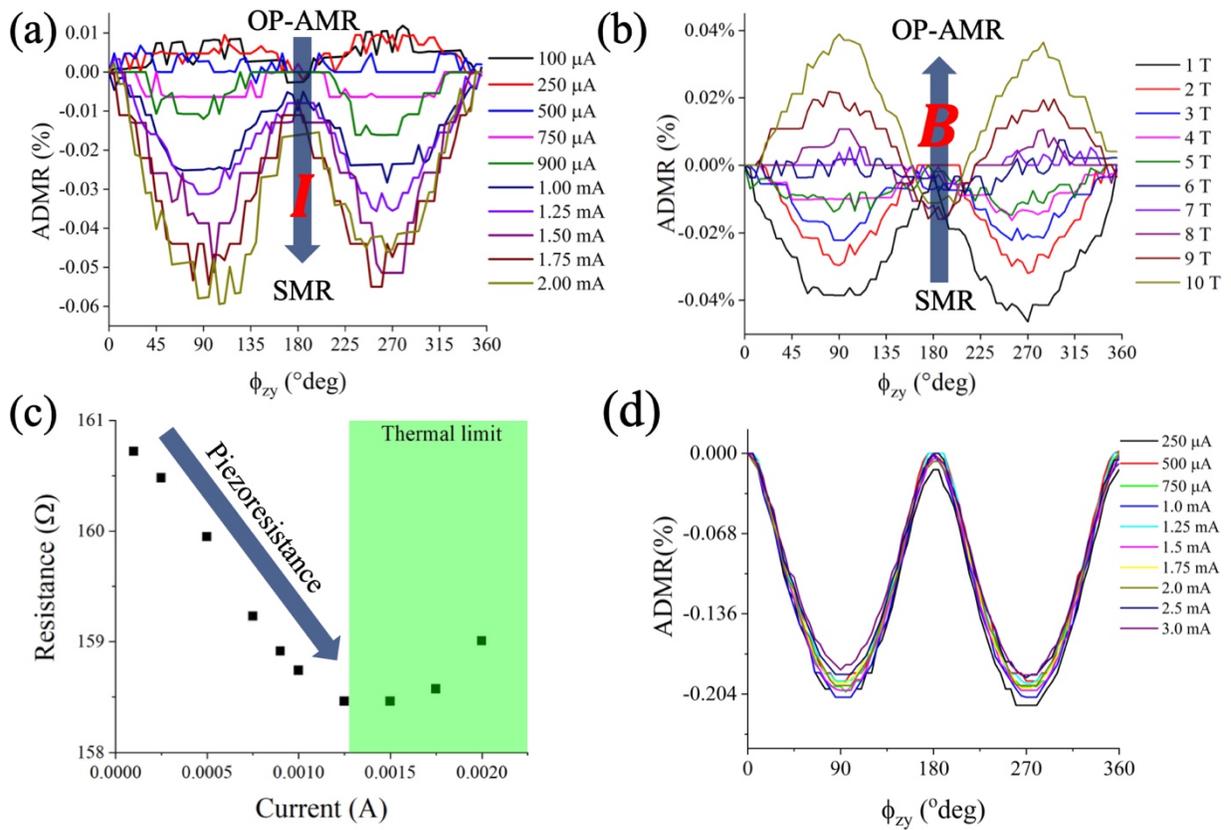

Figure 2. The magneto-transport characterization of Si thin film samples. The angle dependent magnetoresistance as a function of rotation in zy-plane for Pd/Ni$_{80}$Fe$_{20}$/MgO/p-Si thin film system aligned along <110> direction (a) as a function of current showing transition from weak OP-AMR

behavior to SMR at higher currents and (b) as a function of magnetic field showing transition from SMR at low fields to AMR at higher fields. Large arrow indicates direction of current and magnetic field change. (c) the electrical resistance as a function of current at 300 K showing decrease attributed to piezoresistance in p-Si along <110> increasing compressive strain and (d) The angle dependent magnetoresistance as a function of rotation in zy-plane for Pd/Ni$_{80}$Fe$_{20}$/MgO/n-Si thin film system aligned along <110> direction showing SMR behavior as a function of current for an applied magnetic field of 3 T.

### C. SMR measurement in n-Si thin film

We attribute the SMR behavior in p-Si to spatially varying strain. Hence, strain effects should be observed in n-Si specimens as well. To verify it, we fabricated a Pd (1 nm)/Ni$_{80}$Fe$_{20}$ (75 nm)/MgO (1.8 nm)/n-Si (2 μm) freestanding structure in the same 4-probe measurement geometry as shown in Fig. 1 (b,c). The thickness of the Ni$_{80}$Fe$_{20}$ layer is increased, since the n-Si is more conductive ($\rho_{n-Si} = 1.94 \times 10^{-5}$ Ωm) than the p-Si. Thicker Ni$_{80}$Fe$_{20}$ thin films do not exhibit OP-AMR behavior leading to simplified SMR estimates. The n-Si layer in the free-standing structure is oriented along Si <110> similar to p-Si. ADMR measurements as a function of current from 250 μA to 3 mA were taken as shown in Supplementary Figure S11. The resistance of the specimen increases with increasing current contrary to that of the p-Si sample. The sign of the piezoresistance coefficient for n-Si is opposite to that of p-Si. Hence, the compressive stress leads to an increase in resistance in the n-Si due to the piezoresistive effect. The measured SMR is 0.192% and shows a small decrease when the applied current is increased as shown in Figure 2 (d) unlike the p-Si sample. The magnitude of SMR is approximately twice as large as that of p-Si, which itself is larger than that of Pt. Assuming a spin diffusion length of 2 μm[44] for n-Si, we calculate the spin Hall angle to be 0.119 (Supplementary information-H[14]), which is larger than the calculated spin-Hall angle for the p-Si sample. The SMR behavior as a function of current in n-Si suggests that the strain dependent behavior of the n-Si sample is different from p-Si sample. The residual stresses evolve during Si and SiO$_2$ etching and other processing steps. The Ni$_{80}$Fe$_{20}$ layer is three times thicker in case of n-Si and the resulting residual stresses in n-Si after processing could be sufficient for the observation of the SMR behavior. In addition, the thin film structure can buckle to have either a convex or concave curvature, which will give rise to different signs for the strain gradient. However, further change in stresses (strain) due to thermal expansion may be relatively small to cause a significant change in the SMR behavior in case of n-Si.

### D. Crystallography dependent MR in p-Si

The next step is to understand the effect of a ~4% tensile strain near the interface. We performed density functional theory (DFT) calculations of the band structure of Si with an applied strain along <001> and <110> directions. (Supplementary information- I[14] and see, also, references [45-50]). The applied strain lifts the degeneracy of the valence band maxima resulting in a strain mediated valence band splitting. A 4% tensile strain applied along <001> direction leads to an energy splitting of 317 meV in the valence band as shown in Figure 3 (a) and for compressive strain, the splitting increases to 412 meV as shown in Supplementary Figure S12 (d). Similarly, along <110>, an applied 4% tensile or compressive strain leads to valence band splitting of ~520 meV or 600 meV respectively as shown in Figure 3 (b) and Supplementary Figure S12 (a)-(b) respectively. Applied strain has a significantly larger effect on the valence bands than on the conduction bands as shown in Supplementary Figure S12 (c)-(d). The fact that the SMR in p-Si (n-Si) has a strong (weak) dependence on the current is consistent with the picture of the SMR driven by temperature-controlled strain.

From DFT simulations, we observed that the valence band splitting due to strain in the <110> direction is different from that due to strain in the <100> direction. The symmetry of <110> strained Si will be lower than the <100>strained Si[51], which will give rise to a crystallographic dependent behavior. To ascertain the crystallographic direction dependent behavior, we fabricated a set of $Ni_{80}Fe_{20}$ (25 nm) / MgO (1 nm) / p-Si (2μm) multilayer structures with the longitudinal direction of the Si layer lying along <110>, and at 15°, 30° and 45° with respect to <110> as shown in Figure 3 (c). The negative MR for the Si channel oriented along <110> has two kinks due to changes in slope indicated by the arrows. The kink at higher magnetic field (~1.1 T) corresponds to the change in slope at the saturation magnetization ($M_s$). The kink at low field (~0.2 T) is not expected for a $Ni_{80}Fe_{20}$ thin film hard axis magnetization. This kink can only arise due to spin dependent tunneling across the oxide barrier. The low field kink disappears for measurement along <100> direction or at 45 degrees from <110>, which indicates the changes in the spin dependent interactions between the $Ni_{80}Fe_{20}$ and the Si layers. The negative MR behavior arises from polycrystalline $Ni_{80}Fe_{20}$ thin film. And, the observed correlation of the MR with the Si layer crystallographic direction will not arise if there are no spin dependent tunneling and interactions. This measurement gives additional proof that exchange interactions are taking place in spite of thick oxide layer (MgO and $SiO_2$). We have also demonstrated that the spin dependent interactions are function of crystallographic direction of p-Si layer.

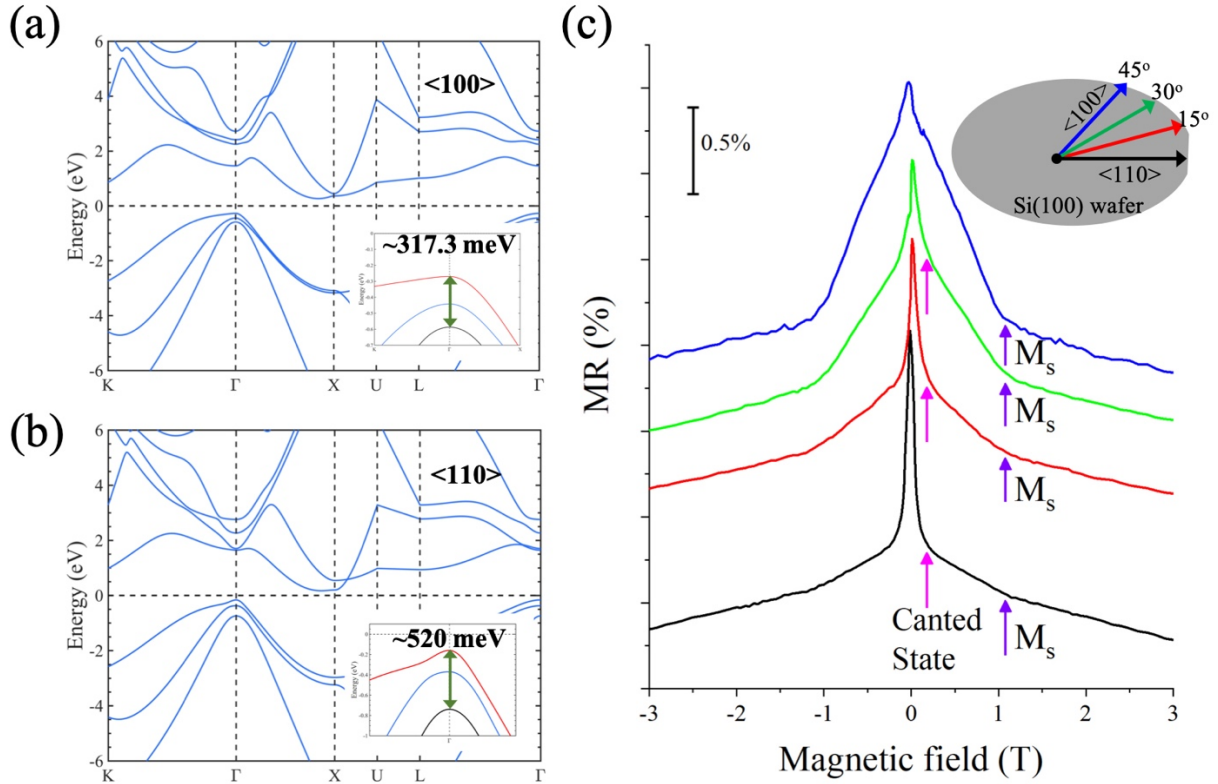

Figure 3. Crystallographic behavior. (a) the band structure of Si for 4% tensile strain applied along <100> and inset showing the energy splitting at the peak of valence band, (b) the band structure of Si for 4% tensile strain applied along <110> and inset showing the energy splitting at the peak of valence band and (c) the magnetoresistance for an applied out of plane magnetic field for current applied along <110> direction or along the flat of the Si(100) wafer, at 15° to the <110> direction, at 30° to the <110> direction and along <100> direction. Arrows showing saturation magnetization and possible canted states and its transition as a function of orientation.

## IV. Discussion

There are various mechanisms that can give rise to SMR behavior in Si. From experimental transport measurements, we demonstrate that inhomogeneous strain is the macroscopic cause of the SMR response in n- and p-doped Si. Microscopically, inhomogeneous strain can be modeled as Rashba-Dresselhaus SOC (Supplementary information-A[14]), which can give rise to SMR behavior. However, the spin-Hall angle is larger than the one expected according to flexoelectric coefficients (Supplementary information-A[14]). Alternatively, inhomogeneous strain will give rise to internal effective magnetic field, which arises due to coupling of electron spin to the off diagonal elements of crystallographic strain tensor as described by Crooker et al.[4] and can give rise to SHE but that can also not explain the large magnitude of SMR observed in this study.

Recently, Lou et al.[52] demonstrated spin-phonon interactions leading to a change in thermal conductivity in both p-Si and n-Si[21,22]. While the charge carriers in p- and n-doped Si are different, the thermal transport is phonon mediated in both cases. We speculate that spin dependent electron-phonon scattering may also give rise to the observed SMR behavior reported earlier. In order to uncover the mechanistic origin of the behavior, we measured the transverse spin-Nernst effect (SNE) in p-Si (Supplementary information- J[14] and see, also, references [53,54]). While the magneto-thermal transport measurement shows transverse SNE behavior in the measurements, experimental results are inconclusive as shown in Supplementary Figure S14. However, these measurements do indicate the existence of interlayer spin-phonon coupling.

In heavy metals, mechanistic reason for both SHE and SNE is large SOC[55]. However, that is not true for Si where thermal transport is mediated by phonons as opposed to charge carriers. Microscopically, spin dependent interactions with phonons cause transverse spin current or SHE during charge transport. And, an inverse microscopic behavior occurs during thermal transport where phonons have spin dependent interactions with charge carrier and give rise to transverse spin current or SNE. Hence, a strain mediated spin dependent coupling between phonon and charge carrier is proposed to be the microscopic mechanism for SHE observed in this study.

**Conclusion**

This study presents an experimental evidence of inhomogeneous strain mediated spin-phonon coupling in centrosymmetric non-magnetic material[56]. The spin-phonon coupling and resulting efficient spin to charge conversion may be applicable to all diamond cubic semiconductors (GaAs, Ge, InSb etc.) under inhomogeneous strain. Manufacturing processes for strain engineering already exist not only for Si but also for other semiconductors. Topological behavior can also arise from the inhomogeneous strain fields, which may also open simple materials systems for topological materials research irrespective of intrinsic spin-orbit coupling. In addition to proposed experimental studies, theoretical models that describe the spin-phonon coupling in centrosymmetric materials and resulting behavior also need to be developed. This work provides a starting point for such future studies.

## Author contributions

PCL, AK and RGB have equal contribution to this work. PCL carried out the SMR and SNE setup fabrication and did the SMR measurements. AK and RGB fabricated the crystallography device fabrication and contributed to all the measurements. TB carried out the DFT calculations. SK conceived the idea, designed the experiments, coordinated the project, supervised the experimental work, performed data analysis, and led manuscript preparation. RKL supervised the theory and computational work and analyzed the experimental data. SK, WPB and RKL wrote the manuscript. All the authors discussed and commented on the manuscript.

## Acknowledgement

The fabrication of experimental devices was done at Center for Nanoscale Science and Engineering at UC Riverside. Electron microscopy sample preparation and imaging was done at the Central Facility for Advanced Microscopy and Microanalysis at UC Riverside. Theory and analysis was supported as part of Spins and Heat in Nanoscale Electronic Systems (SHINES) an Energy Frontier Research Center funded by the U.S. Department of Energy, Office of Science, Basic Energy Sciences under Award No. #DE-SC0012670. This work used the Extreme Science and Engineering Discovery Environment (XSEDE),[50] which is supported by National Science Foundation Grant No. ACI-1548562 and allocation ID TG-DMR130081.

## Data availability

The data that support the findings of this study are available from the corresponding author upon reasonable request.

## Competing interests

The authors declare no competing interests.

**Supplementary information- Large spin-Hall effect in Si at room temperature**

Paul C. Lou[1], Anand Katailiha[1], Ravindra G Bhardwaj[1], Tonmoy Bhowmick[2], W.P. Beyermann[3], Roger K. Lake[2,4] and Sandeep Kumar[1,4*]

[1] Department of Mechanical Engineering, University of California, Riverside, CA 92521, USA

[2] Department of Electrical Engineering and Computer Science, University of California, Riverside, CA 92521, USA

[3] Department of Physics and Astronomy, University of California, Riverside, CA 92521, USA

[4] Materials Science and Engineering Program, University of California, Riverside, CA 92521, USA

**A -Rashba-Dresselhaus spin-orbit coupling (RD-SOC) in centosymmetric material under strain and strain gradient**

To describe the RD-SOC in Si for electronic transport, we model it using a Hamiltonian that can be written approximately as:

$$H = H_0 + H_1 + H_2 + H_3 + H_4 \tag{S1}$$

The first term represents centrosymmetric Si, $H_1$ is the flexoelectric effect mediated bulk Rashba SOC, $H_2$ represents the strain gradient mediated Dresselhaus spin splitting, $H_3$ represents the strain mediated Dresselhaus spin splitting, and $H_4$ represents the shear strain mediated Rashba spin splitting. In the case of Si, the strain gradient is essential for last four terms in Hamiltonian. Strain spin splitting exists only in non-centrosymmetric semiconductors. A strain gradient breaks the inversion symmetry of the Si crystal structure. The flexoelectric response can written as:

$$E_l = \frac{\mu_{ikjl}}{\varepsilon} \frac{\partial \epsilon_{ij}}{\partial x_k} \tag{S2}$$

Where $E$, $\mu$, $\varepsilon$ and $\epsilon$ are electric field, flexoelectric coefficient, dielectric constant, and strain, respectively. The flexoelectric polarization will give rise to interfacial Rashba spin-orbit coupling[1] written as:

$$H_R \propto (\vec{E} \times \vec{p}) \cdot \vec{\sigma} = E_3(-p_y\sigma_x + p_x\sigma_y) = \frac{\mu_{xx,zz}}{\varepsilon} \frac{\partial \epsilon_{xx}}{\partial x_z}(-p_y\sigma_x + p_x\sigma_y) \tag{S3}$$

Where $p$ and $\sigma$ are angular momentum and spin polarization respectively. A centosymmetric Si lattice is proposed to have site inversion asymmetry, which results in hidden Dresselhaus (D2 type)[2] spin splitting. Hence, strain gradient mediated bulk inversion asymmetry (BIA) will also contribute towards Dresselhaus spin splitting. In addition, a second Dresselhaus type spin splitting will arise due to the normal strain in the presence of strain gradient mediated BIA. The resulting Hamiltonian can be described as

$$H = H_0 + \frac{\mu_{xx,zz}}{\varepsilon}\frac{\partial \epsilon_{xx}}{\partial x_z}(-p_y\sigma_x + p_x\sigma_y) + D_1\left(\frac{\partial \epsilon_{xx}}{\partial x_z}\right)(p_x\sigma_x - p_y\sigma_y) + D_2(-\epsilon_{xx})(p_x\sigma_x - p_y\sigma_y) + D_3\epsilon_{xz}(-p_y\sigma_x + p_x\sigma_y), \tag{S4}$$

where $D_1$, $D_2$ and $D_3$ [3,4] are material parameters. The resulting Hamiltonian can be written as

$$H = H_0 + \beta(-p_y\sigma_x + p_x\sigma_y) + \lambda(p_x\sigma_x - p_y\sigma_y), \tag{S5}$$

where $\beta$ and $\lambda$ are Rashba and Dresselhaus parameters respectively. The total Hamiltonian is then

$$H = \frac{\hbar^2 K^2}{2m_{hh}} + H_R + H_D . \tag{S6}$$

In this study, the premise of this hypothesis depends on the flexoelectric polarization of Si. Recently, Schiaffino et al. reported the flexoelectric coefficients for Si ($\mu_{1111} = -1.411\ nC/m$, $\mu_{1122} = -1.049\ nC/m$, and $\mu_{1212} = -0.189\ nC/m$) using a metric wave approach[5]. Using these flexoelectric coefficients and for a strain gradient of $4 \times 10^4$ m$^{-1}$, we estimate a spontaneous polarization of $4.196 \times 10^{-5}$ C/m$^2$, which is an order of magnitude smaller than the spontaneous polarization of GaAs[6]. However, the flexoelectric effect can be an order of magnitude larger near the interface and at nanoscale[7-9] and may give rise to SHE behavior having magnitude similar to GaAs. Instead, the experimental values suggest a SHE two order of magnitude larger than GaAs.

**B- Materials and methods**

Device fabrication: We choose a commercially available 4" silicon-on-insulator (SOI) wafer with 2 μm thick device layer (resistivity 0.001- 0.005 Ω-cm), 1 μm oxide layer and 300 μm handle layer (resistivity 1-20 Ω-cm). Using UV photolithography and deep reactive ion etching (DRIE), the handle layer (back side) is etched underneath the specimen region as seen in the Figure 1(c), followed by patterning and etching the Si specimen on the device layer (front side). The silicon structure is made freestanding by etching the sacrificial oxide (box layer of SOI wafer) using hydrofluoric (HF) acid vapor etching. The surface oxide is removed using Ar milling for 15 minutes followed by a layer of 1.8 nm of MgO using RF sputtering. A layer of 25 nm $Ni_{80}Fe_{20}$/ 1 nm Pd is deposited onto the device using e-beam evaporation. The MgO layer is required for efficient spin tunneling and diffusion inhibitor whereas the Pd layer inhibits the oxidation of the $Ni_{80}Fe_{20}$ layer. Similar method is used to fabricate the n-Si specimen and crystallography specimens. The 1 μm oxide layer electrical isolate the electrodes. P-Si is doped with Boron and n-Si with Phosphorous.

SMR measurement: The experiments for SMR measurements are performed in Quantum Design's Physical Properties Measurement System (PPMS). The transport properties are measured using an alternating current (AC) lock-in technique using Keithley 6221 current source and Stanford Research Systems SR830 lock-in amplifier.

MR measurement: The MR measurement on the p-Si sample is carried out at 900 μA of current.

Magnetic characterization: The magnetic characterization is carried out inside a Quantum Design magnetic property measurement system (MPMS) at 208 K. The field is swept from 1525 Oe to -1525 Oe. The sample layered structure is same as the SMR p-Si sample.

TEM sample preparation: We deposited MgO (1.8 nm), $Ni_{80}Fe_{20}$ (25 nm) and MgO (50 nm) layers on top of this control Si thin film specimen. The additional MgO layer is to protect the specimen during focused ion beam (FIB) sectioning for microstructure and interfacial study. TEM lamellae were prepared from the layered sample following established procedures with a DualBeam scanning electron microscope and FIB instrument using Ga ion source (Quanta 200i 3D, ThermoFisher Scientific). First, a strap of 5 μm thick protective Carbon layer was deposited over a region of interest using the ion beam. Subsequently approximately 80 nm thin lamella of was cut and polished at 30 kV and attached to a TEM grid using in-situ Omniprobe manipulator. To reduce surface amorphization and Gallium implantation final milling at 5 kV and 0.5 nA was used to thin the sample further.

S/TEM imaging and analysis: TEM and STEM imaging was performed at 300 kV accelerating voltage in a Thermo Fisher Scientific Titan Themis 300 instrument, fitted with X-FEG electron source, 3 lens condenser system and S-Twin objective lens. High-resolution TEM images were recorded at resolution of 2048x2048 pixels with a FEI CETA-16M CMOS digital camera with beam convergence semi-angle of about 0.08 mrad. STEM images were recorded with Fischione Instruments Inc. Model 3000 High Angle Annular Dark Field (HAADF) Detector with probe current of 150 pA, frame size of 2048x2048, dwell time of 15 μsec/pixel, and camera length of 245 mm.

SNE sample fabrication: The fabrication process for the SNE device is same as SMR device except that back side DRIE is not done. The specimen is made freestanding using longer HF vapor etch.

## C- Strain estimation

The length of primary SMR specimen is 160 µm and for this length the deflection due to buckling is extremely small. To overcome this problem, we make another sample having length of 600 µm to measure the residual stresses. Once the specimen is made freestanding, the control specimen buckles due to residual stress in the Si even before deposition of MgO and $Ni_{80}Fe_{20}$ thin film layers as shown in Supplementary Figure S1 (b). This buckling of control beam clearly corroborates the existence of a strain gradient. The deflection is measured to be 8.73 µm for length 600 µm. The buckling in the plane of thin film requires large stresses (~>4 GPa). The Si device layer in Si on insulator (SOI) wafers do not have such large residual stresses. The in-plane buckling arises due to the HF vapor etching process used to make the specimen freestanding. As the oxide is etched laterally (along width), the stress relaxation leads to in-plane buckling since thin oxide layer (along the width direction) does not allow out of plane deformation. We fabricate another control specimen having the same dimensions as the SMR specimen shown in Figure S1 (b), which also bends when made freestanding. These beams show the bending of the beam, which is the origin of strain gradient.

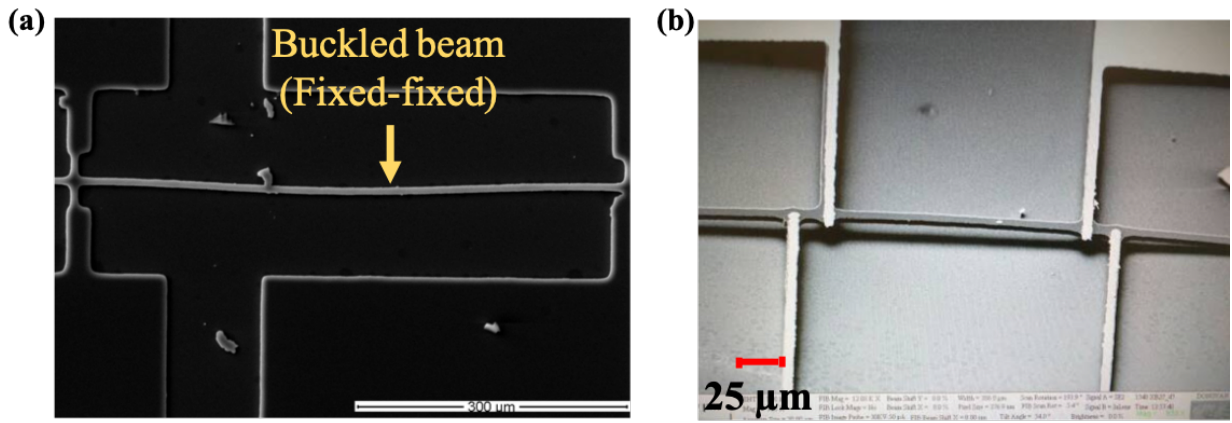

Supplementary Figure S1. (a) The buckling of control freestanding Si beam (having the same layered structure as the p-Si SMR sample in the main text) due to residual stresses, and (b) buckling of the Si only beam having same length and structure as specimen used for SMR study.

We deposit MgO and $Ni_{80}Fe_{20}$ thin film layers to replicate the SMR specimen heterostructure. The high-resolution transmission electron microscope (HRTEM) is used to study the interfaces and strain in the thin film. The HRTEM sample is cut from the center of Si freestanding beam shown in Supplementary Figure S1 (a). Using HRTEM, we estimate the tensile strain to be 4% near the interface. To estimate strain, we plot the intensity profile along <110> direction. Using Gaussian fit, each peak is identified followed by peak to peak distance and average over multiple measurements is carried out as shown in Supplementary Figure S2. This analysis is carried out near the interface and away from the interface.

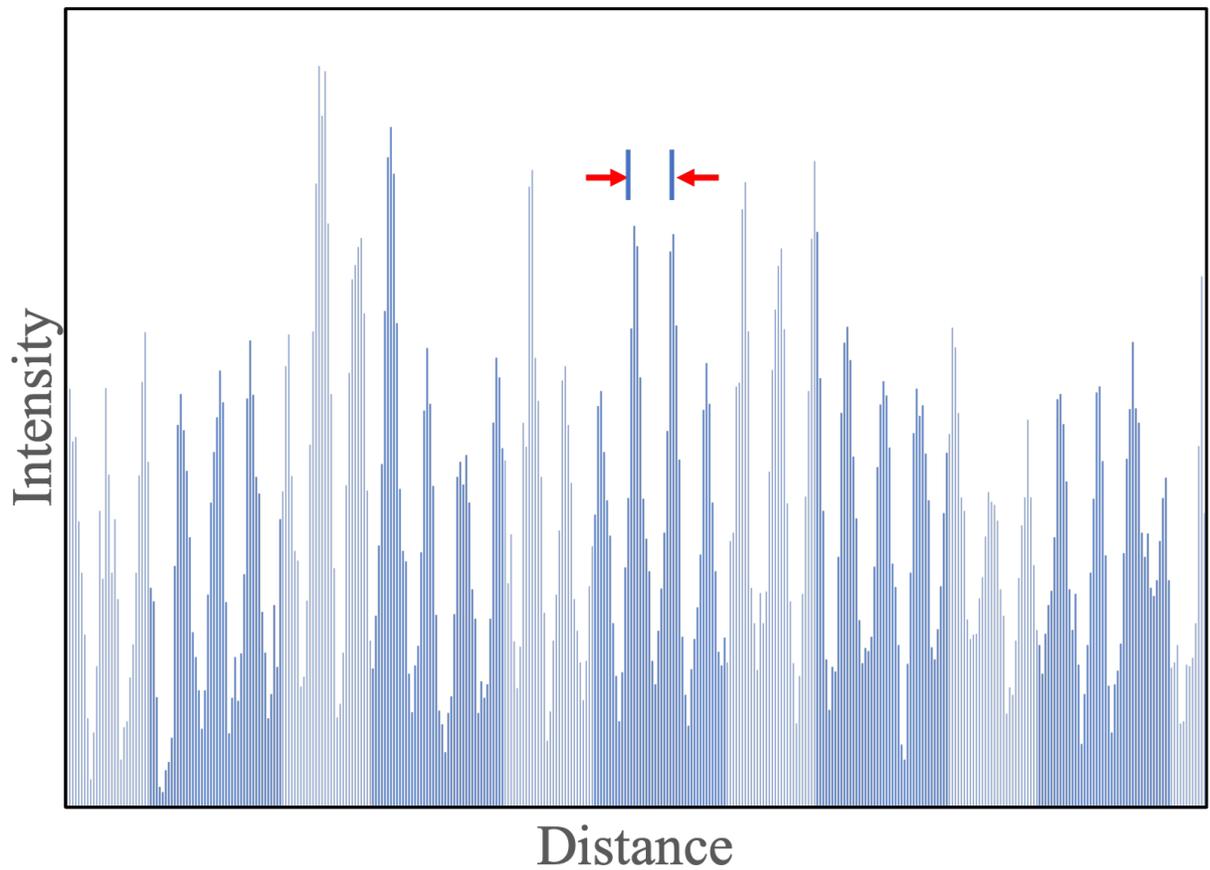

Supplementary Figure S2. The intensity profile along <110> direction used to identify the strain.

**D- Elemental mapping**

To verify Ni or Fe diffusion into Si layer, energy-dispersive X-ray Spectroscopy (EDS) analysis and elemental mapping were obtained in the STEM at 300 kV, utilizing ThermoFisher Scientific SuperX system equipped with 4x30mm$^2$ window-less SDD detectors symmetrically surrounding the specimen with a total collection angle of 0.68 srad, by scanning the thin foil specimens. Elemental mapping was performed with an electron beam probe current of 550 pA at 1024 x1024 frame resolution. And the resulting data does not show existence of any measurable Ni or Fe diffusion in Si layer as shown in Supplementary Figure S2 (a,b). Elemental map of O corresponds to SiO$_2$ insulator layer that isolates the device Si layer from handle Si layer as shown in supplementary Figure S2 (a).

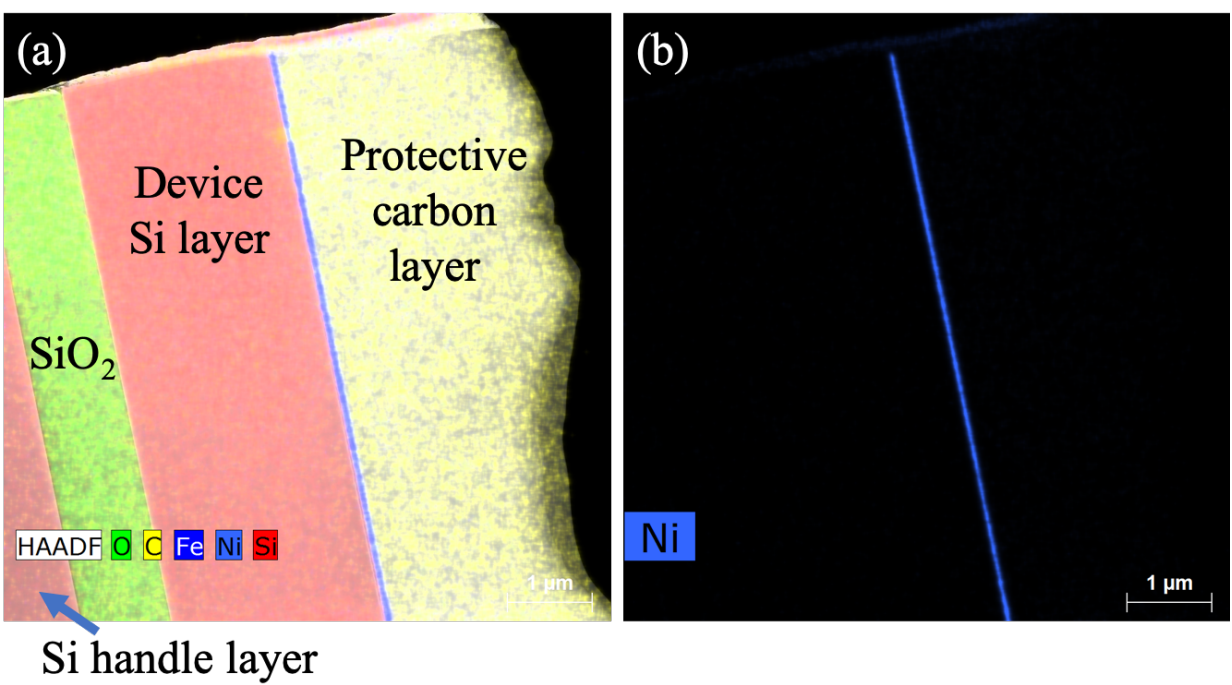

Supplementary Figure S3. The EDS elemental mapping showing (a) the elemental distribution of Si, O, Ni, Fe and C across the multilayer heterostructure and (b) elemental distribution of Ni. The C layer is used for protection during FIB sample preparation as described in materials and methods.

**E- Spin dependent interactions across $Ni_{80}Fe_{20}$ and p-Si**

As stated in the main text, the 5.5 nm thick oxide layers, observed in HRTEM measurement, can quench any spin tunneling and SMR behavior. To ascertain the spin tunneling and exchange interactions, we measured MR for an applied magnetic field from 14 T to -14 T along y- and z-axis of the sample. The in-plane (y-axis) MR behavior at low field shows two AMR peaks at 100 Oe and -100 Oe corresponding to the coercivity of the sample for each magnetic field sweep direction as shown in Supplementary Figure S4.

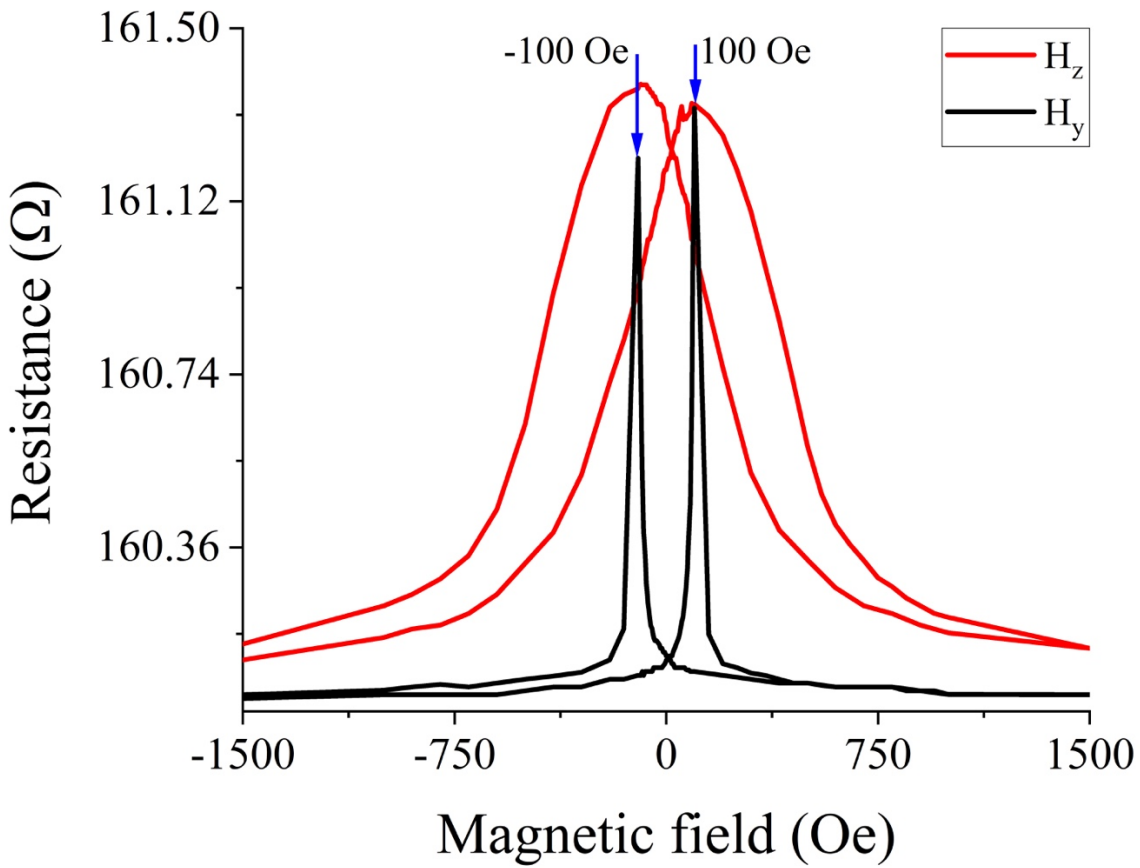

Supplementary Figure S4. The low field magnetoresistance behavior of the p-Si heterostructure sample showing AMR peak at 100 Oe, which arises due to spin-valve behavior from spin dependent interactions.

However, the coercive field for $Ni_{80}Fe_{20}$ thin film is expected to be ~10 Oe or smaller[10]. We measured the magnetic hysteresis at 208 K for a layered thin structure similar to our transport measurement sample as shown in Supplementary Figure S5, which clearly shows a coercive field significantly smaller than 100 Oe. The magnetic hysteresis is similar at 300 K as well [11]. Based on MR and magnetic hysteresis measurement, the AMR peaks can only arise due to spin dependent tunneling and exchange interactions leading to spin valve behavior in spite of thick oxide layers.

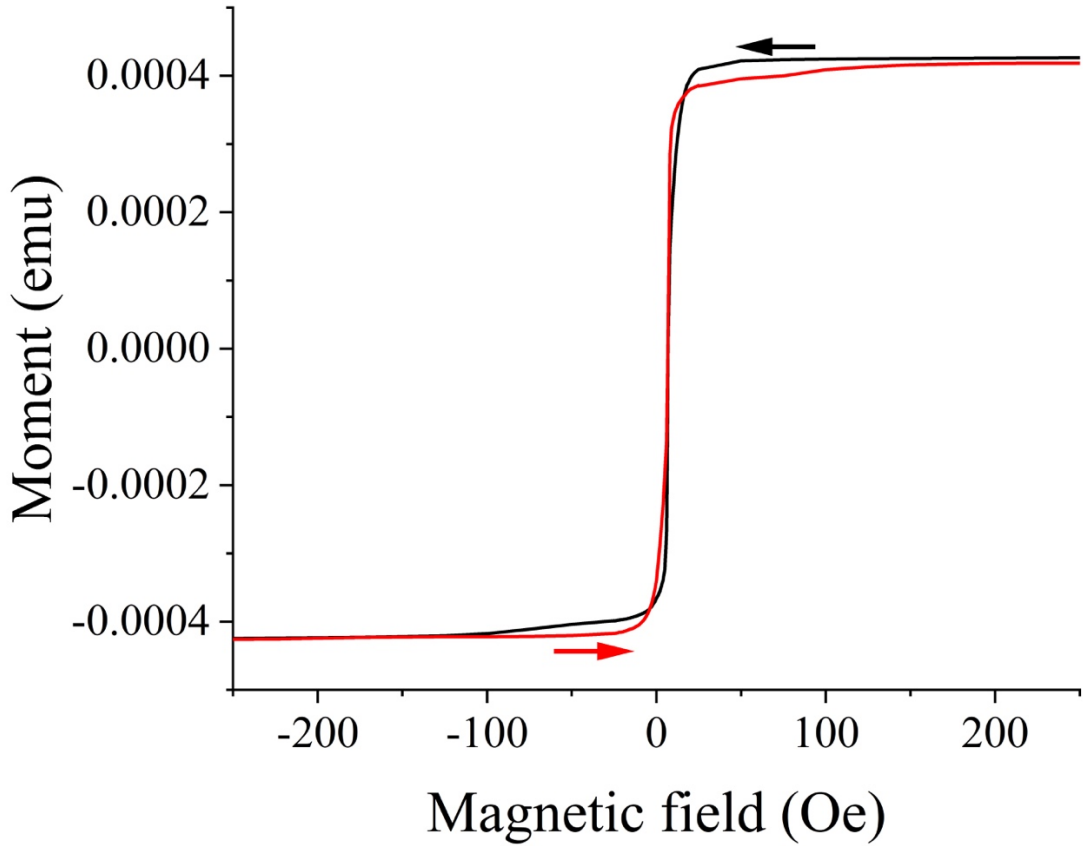

Supplementary Figure S5. The magnetic hysteresis of $Ni_{80}Fe_{20}$ and p-Si heterostructure thin film sample(unetched) at 208 K showing coercivity smaller than 100 Oe.

We, then, analyzed the MR behavior for whole magnetic field range from 14 T to -14 T. As shown in Figure 2 (b), a transition from SMR to OP-AMR behavior is observed around 6 T. We observe a similar crossover behavior between in-plane and out of plane MR as shown in Supplementary Figure S6. The AMR in ferromagnetic thin films arises due to relative orientation of current direction and magnetization. In addition to AMR, the ferromagnetic thin films also exhibit a negative MR for fields larger than saturation magnetic field, which is attributed to reduction in magnon-electron scattering at high fields. The negative MR due to magnons has same slope at higher fields irrespective of in-plane or out of plane magnetic field[12]. This can be verified from OP-AMR measurement for $Ni_{80}Fe_{20}$ thin film shown in Supplementary Figure S7, where magnitude of OP-AMR is same for both 4T and 8T. However, in case of composite sample, the slope of in-plane MR is smaller than the out of plane MR leading to a crossover and diverging behavior. This increase in slope for out of plane MR is attributed to the spin current absorption in $Ni_{80}Fe_{20}$ layer due to the SHE in p-Si layer. The ADMR and MR measurements in p-Si sample presented so far clearly support the observation of SMR behavior due to spin dependent interactions.

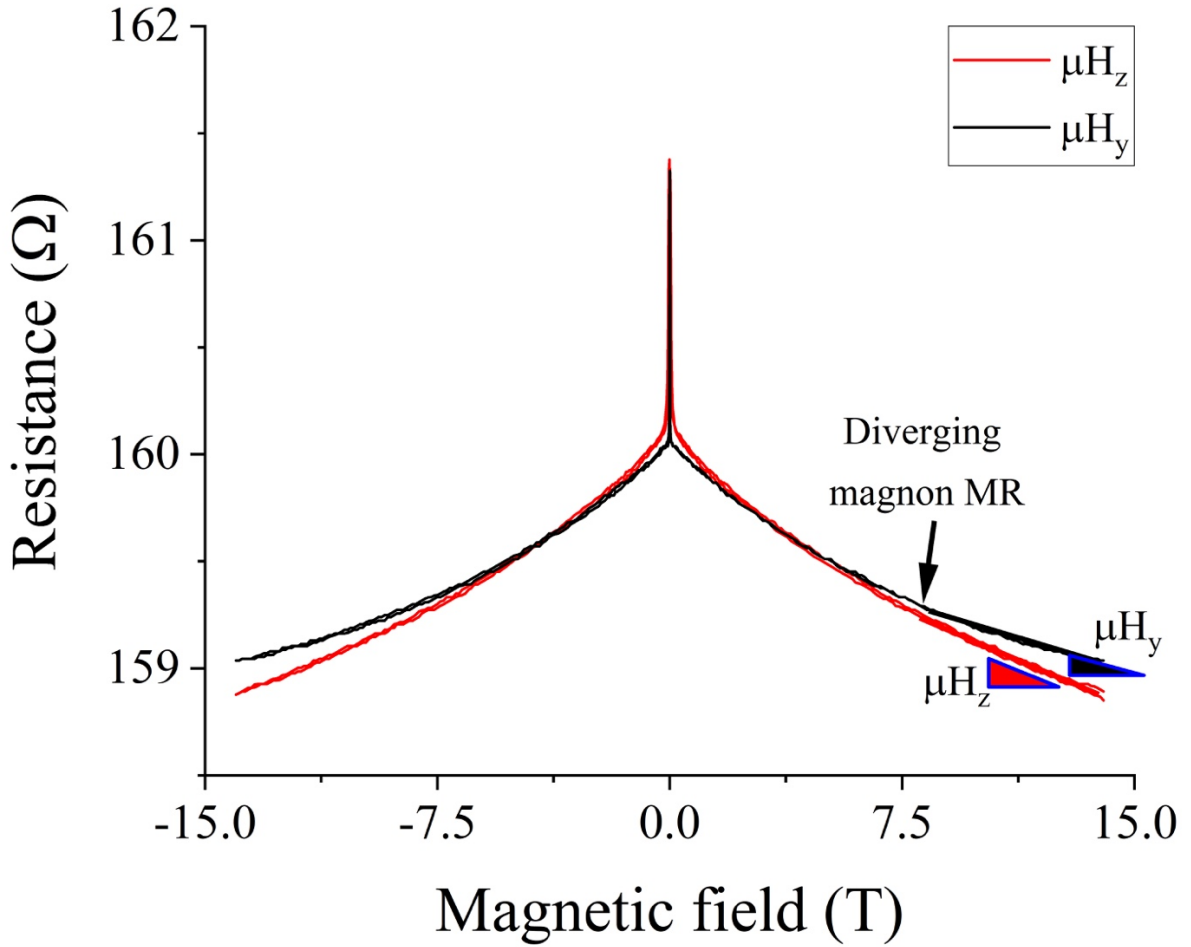

Supplementary Figure S6. The high field magnetoresistance behavior of the p-Si heterostructure sample showing diverging magnon MR behavior due to spin dependent interactions.

# F- OP-AMR measurement in 25 nm thick $Ni_{80}Fe_{20}$ thin film

We fabricated first control sample of $Ni_{80}Fe_{20}$ (25 nm) on a thermal oxide (300 nm) Silicon wafer. This sample will allow us to estimate the OP-AMR in the $Ni_{80}Fe_{20}$ thin film as shown in Supplementary Figure S7.

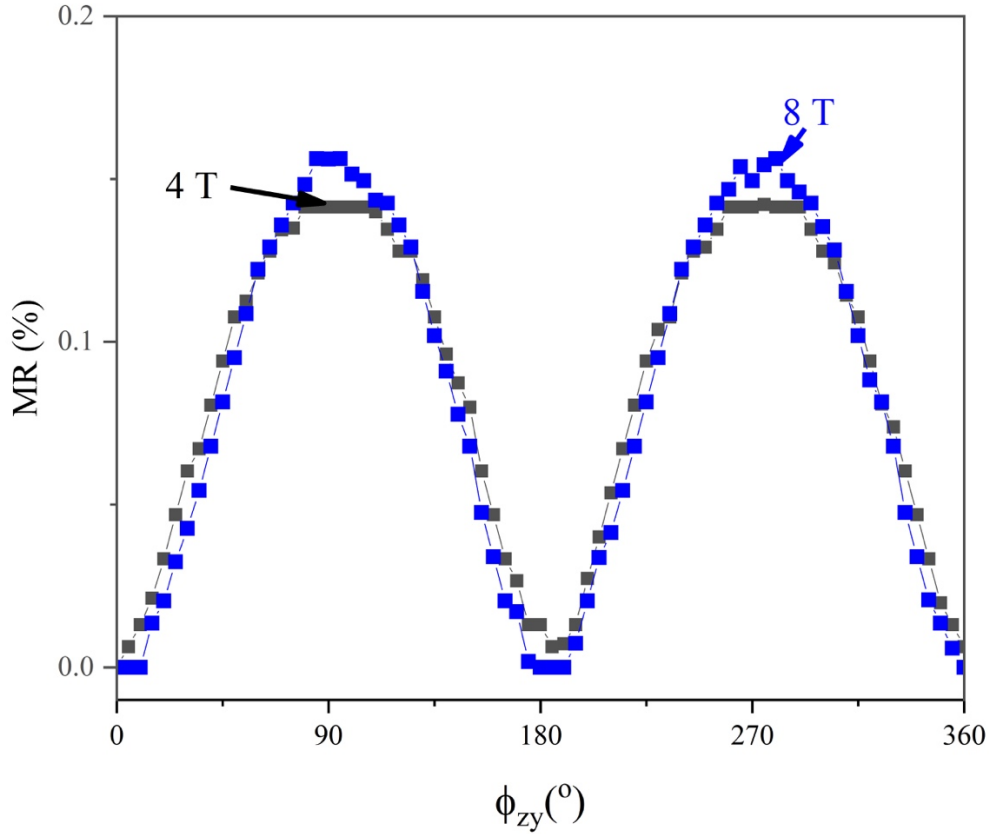

Supplementary Figure S7. The magnetoresistance ratio $\Delta R/R(0°)$ of a 25 nm $Ni_{80}Fe_{20}$ thin film as a function of angular rotation of the magnetic field in the zy-plane at 300 K.

We fabricated a second control sample ($l$=100 μm, $w$= ~15μm) with the following layered structure- $Ni_{80}Fe_{20}$ (25 nm)/$SiO_2$ (Evaporation) (25 nm)/p-Si (2 μm). This sample is freestanding similar to the SMR and SNE samples in the main text. We measured the angle dependent magnetoresistance as a function of magnetic field of 4T in the zy-rotational plane as shown in Supplementary Figure S8. The thick oxide will not allow spin tunneling and the SMR behavior will disappear as shown in this figure. The applied magnetic field is 4 T and multiple (three) measurements show the repeatability of the data. The value of OP-AMR is smaller than that of the $Ni_{80}Fe_{20}$ only sample shown in Supplementary Figure S7. The intermediate oxide layer in this control sample is deposited using e-beam evaporation, which will have pin holes and higher roughness as compared to the thermally grown oxide wafer used for Supplementary Figure S7. We cannot have thermal oxide in this case because it will get etched during the HF vapor etching process used to make the sample freestanding.

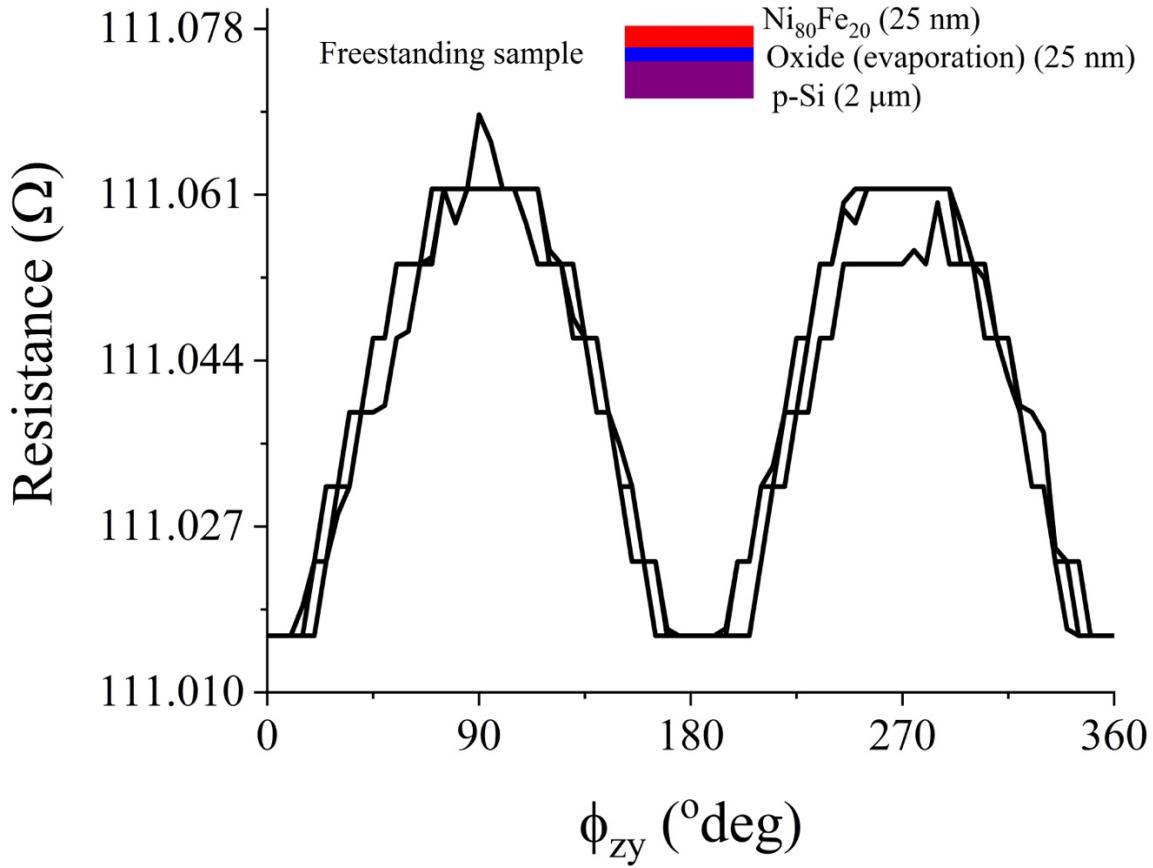

Supplementary Figure S8. The angle dependent magnetoresistance in the zy-plane showing out of plane anisotropic magnetoresistance (OP-AMR) for a control sample with 25 nm of oxide (deposited using e-beam evaporation) in between the $Ni_{80}Fe_{20}$ and p-Si layers.

Strain in the $Ni_{80}Fe_{20}$ layer may also lead to the SMR behavior observed in this study. To eliminate such a possibility, we fabricated a third control sample where $Ni_{80}Fe_{20}$ (25 nm) is deposited on a freestanding oxide membrane. The angle dependent magnetoresistance measurement is done at 1 T, 4 T and 8 T as shown in Supplementary Figure S9. This sample will have large residual stresses since the freestanding oxide membrane will bend and induce strain in the sample. In spite of it, OP-AMR behavior is observed. The resistances of these samples are higher than those of the $Ni_{80}Fe_{20}$ samples on Si due to severe oxidation.

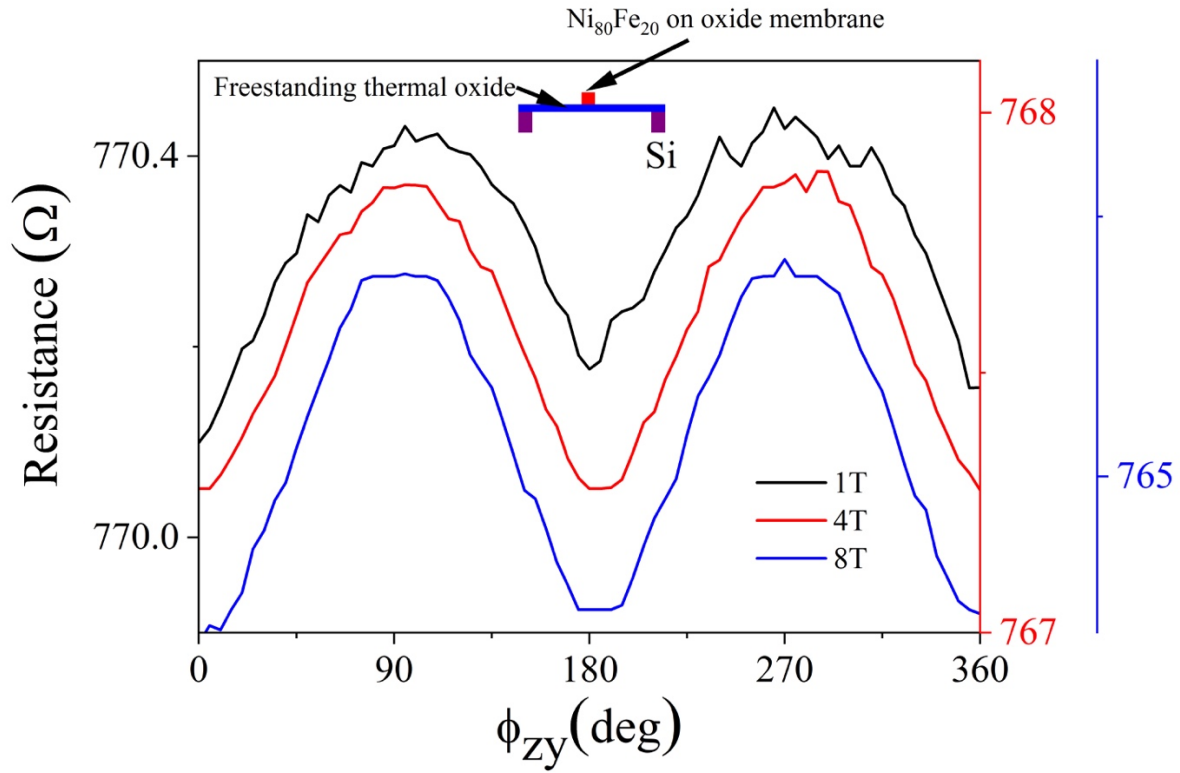

Supplementary Figure S9. Angle dependent magnetoresistance in the zy-plane showing OP-AMR behavior for a $Ni_{80}Fe_{20}$ sample supported on a thermal oxide membrane with no Si underneath.

**G- Resistance as a function of temperature and additional SMR measurement**

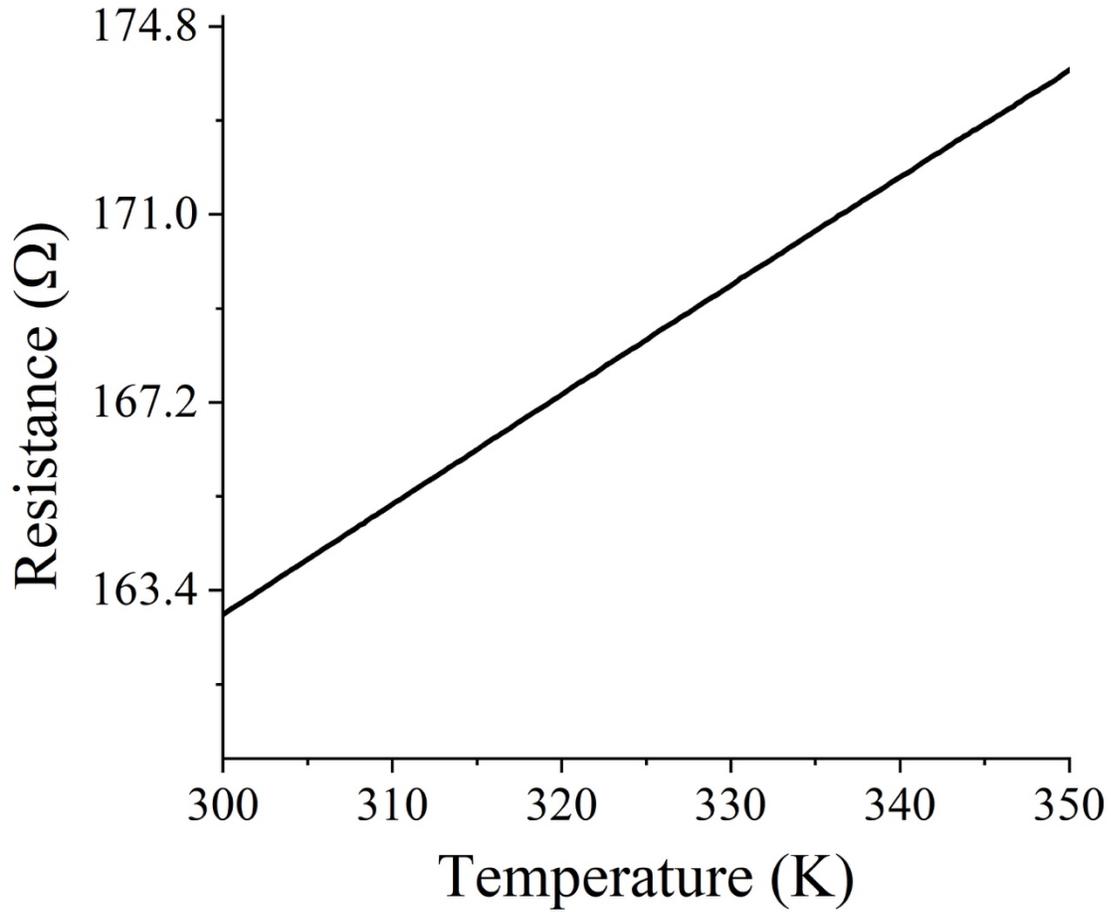

Supplementary Figure S10. The longitudinal resistance of the p-Si SMR specimen as a function of temperature from 300 K to 350 K. It clearly shows that the resistance should increase due to temperature rise.

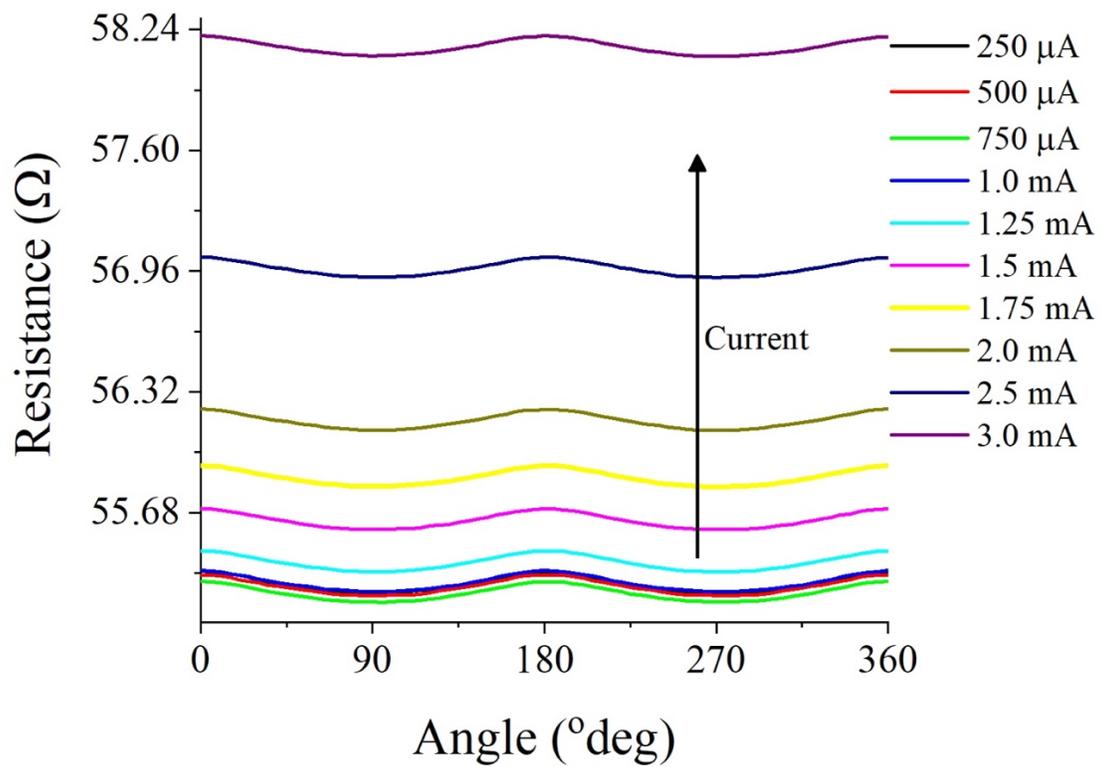

Supplementary Figure S11. The SMR measurement as a function of applied electric current for n-Si specimen. Figure shows increase in resistance as a function of current due to piezoresistance and Joule heating.

## H-Spin-Hall angle estimation

The spin-Hall angle magnetoresistance equation for bimetallic[13] specimen is given as-

$$\frac{\Delta R_{xx}^{SMR}}{R_{xx}^0} \sim - \Theta_{SH}^2 \frac{\lambda_N}{d} \frac{\tanh^2\left(\frac{d}{2\lambda_N}\right)}{1+\xi} \left[\frac{g_R}{1+g_R \coth\left(\frac{d}{\lambda_N}\right)} + \frac{g_F}{1+g_F \coth\left(\frac{d}{\lambda_N}\right)}\right] \quad (S7)$$

where $g_R \equiv 2\rho_N \lambda_N Re[G_{MIX}]$ and $g_F \equiv \frac{(1-P^2)\rho_N \lambda_N}{\rho_F \lambda_F \coth\left(\frac{t_F}{\lambda_F}\right)}$.

In the equation, N represents normal metal (Si in this study) and F-the ferromagnetic metal. We use the following values to calculate the SMR[14]: $\rho_N = 5.25 \times 10^{-5}$ $\Omega$m, $\lambda_N = 310$ nm, $Re[G_{MIX}] = 10^{19} \Omega^{-1} m^{-2}$, $P = 0.7$, $\rho_F = 5.43 \times 10^{-7} \Omega m$, $\lambda_F = 4$ nm, $t_F = 25$ nm, $d = 2$ μm and $\xi = \frac{\rho_N t_F}{\rho_F d} = 1.21$.

With these values, $1 \ll g_R \coth\left(\frac{d}{\lambda_N}\right)$ and $1 \ll g_F \coth\left(\frac{d}{\lambda_N}\right)$. This simplifies the relationship to:

$$\frac{\Delta R_{xx}^{SMR}}{R_{xx}^0} \sim - \Theta_{SH}^2 \frac{\lambda_N}{d} \frac{2*\tanh^2\left(\frac{d}{2\lambda_N}\right)}{(1+\xi)\coth\left(\frac{d}{\lambda_N}\right)} \quad (S8)$$

For $\frac{\Delta R_{xx}^{SMR}}{R_{xx}^0} = 7.88 \times 10^{-4} - 0.00115$, we calculate $\Theta_{SH} \sim 0.075$-$0.096$.

This calculation is highly dependent on the value of spin diffusion length. Shikoh et al.[15] reported spin diffusion length of 148 nm whereas Weng et al. [16] reported spin diffusion length of 40 nm. The resulting spin-Hall angle will be 0.109-0.26. Hence, the spin-Hall angle value for p-Si can be 0.075-0.26.

The calculation is qualitative since the thickness dependent data is unavailable. In addition, the interfacial properties and spin diffusion behavior will be significantly different for strained Si as compared to bulk properties used in the calculations.

For n-Si, using $\frac{\Delta R_{xx}^{SMR}}{R_{xx}^0} = 0.00193$, $\lambda_N = 2$ $\mu m$[17], $\rho_N = 1.94 \times 10^{-5}$ $\Omega m$ and $\xi = \frac{\rho_N t_F}{\rho_F d} = 1.39$, we obtain $\Theta_{SH} \sim 0.118$.

**I-Effect of strain on Si band structure**

To observe the effect of applied strain on Si band structure, ab initio calculations of bulk Silicon are carried out using density functional theory (DFT) with a projector augmented wave method[18] and the Perdew-Burke-Ernzerhof (PBE) type generalized gradient approximation[19,20], as implemented in the Vienna ab initio simulation package (VASP)[21,22]. The Monkhorst-Pack [23] scheme is used for the integration of the Brillouin zone with a k-mesh of 14 Å~ 14 Å~ 14 for the bulk structures. The energy cutoff of the plane wave basis is 300 eV. All of the electronic band structure calculations include spin-orbit coupling. Supplementary Figure S5 shows the additional results.

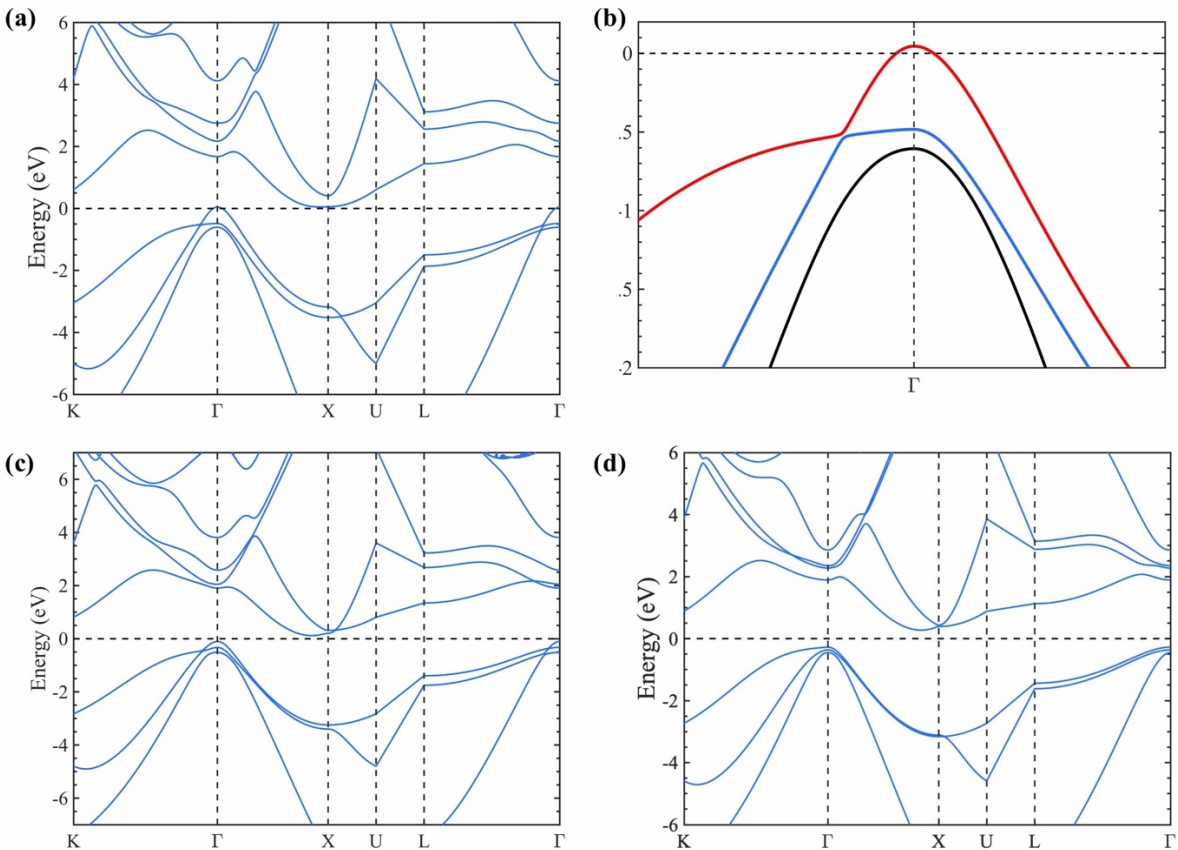

Supplementary Figure S12. Band structure of Si under an applied strain (a) 4% compressive strain along <110>, (b) the valence band maxima at 4% compressive strain along <110> direction, (c) 2% compressive strain along <100>, and (d) 4% compressive strain along <100>.

## J- The transverse spin-Nernst effect measurement

The magneto-thermal transport characterization setup shown in the main text has four Hall junctions. An approximate thermal analysis using COMSOL shows the temperature distribution across the longitudinal direction as shown in Supplementary Figure S13. This analysis shows that the temperature drop is largest for J2 and gradually reduces for J4. Hence, the largest PNE response should be at J2.

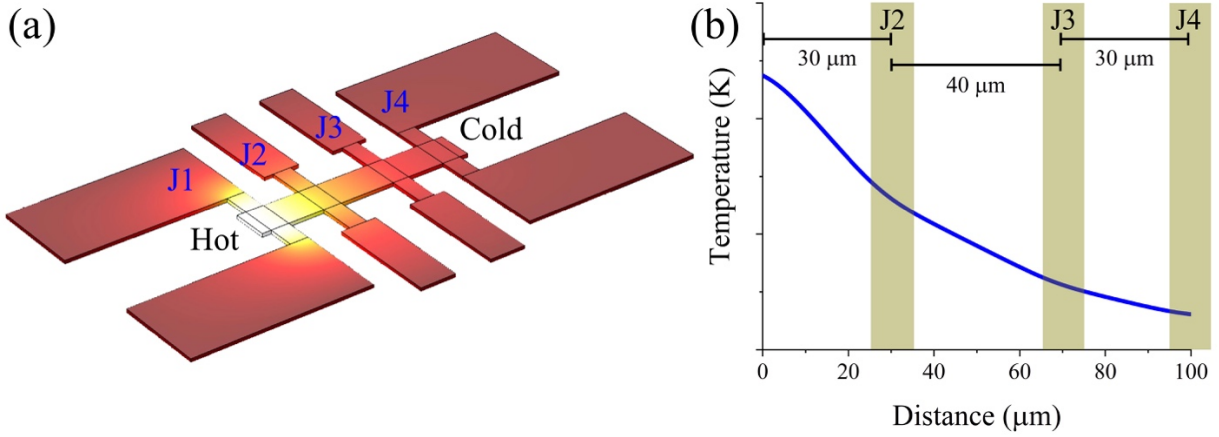

Supplementary Figure S13. (a) the COMSOL simulation of approximate heating effects and resulting temperature gradient in the device structure and (b) the temperature profile along the length of the sample.

In the transverse configuration, the SNE can be written as:

$$E_{SNE} \propto \frac{1}{2}\Delta\alpha_{TSNE}\nabla T_x \sin 2\theta \qquad (S9)$$

Where $\Delta\alpha_{SNE}$ and $\nabla T_x$ are transverse spin-Nernst magneto-thermopower and temperature gradient along the longitudinal direction (x-axis) respectively. Similarly, the planar Nernst effect (PNE) response from the ferromagnetic layer can be written as:

$$E_{PNE} \propto \frac{1}{2}\Delta\alpha_{PNE}\nabla T_x \sin 2\theta \qquad (S10)$$

For an angle dependent measurement in the plane of the structure, both SNE and PNE show a symmetry of $\sin 2\theta$. The resulting response will be combination of these two responses.

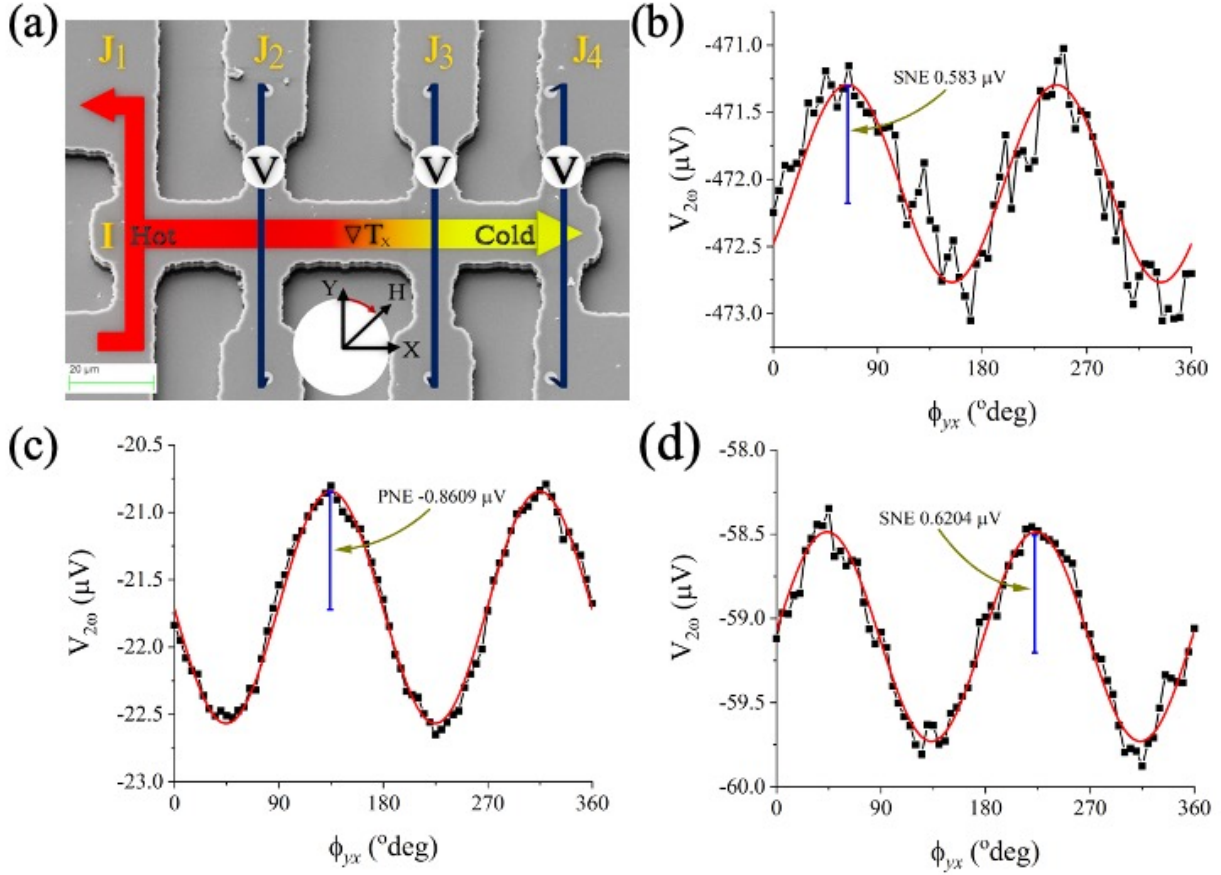

Supplementary Figure S14. The magneto-thermal transport characterization of p-Si thin film sample. (a) a representative scanning electron microscope image showing the schematic of experimental setup and the angle dependent magneto thermal transport measurement in *yx*-plane for an applied magnetic field of 1 T at Hall junctions (b) J2, (c) J3 and (d) J4 showing SNE, PNE and SNE responses respectively. Red line is curve fit.

For the experimental study, we fabricated a setup having multiple Hall bars having p-Si based thin film heterostructure sample as shown in Supplementary Figure S14 (a). In this setup, we applied a heating current across junction J1 and measured the transverse thermal response ($V_{2\omega}$ response) across three Hall junctions J2, J3 and J4 as shown in Supplementary Figure S14 (a). An angle dependent magneto thermal transport measurement at an applied magnetic field of 1 T and 2 mA of heating current is carried out in *yx*-plane for a temperature gradient along the positive *x*-axis. In the absence of any spin dependent response from p-Si, the thermal measurement will show planar Nernst effect (PNE)[24,25] only. PNE and transverse spin-Nernst effect (SNE) have same symmetry ($\sin 2\theta$). However, they will have a phase offset of 90º depending upon the spin-Nernst angle. Our measurement shows a clear PNE response[24] for measurement across junction J3 ($V_{PNE} = -0.861\ \mu V$) as shown in Supplementary Figure S14 (c). However, measurement across junctions J2 ($V_{SNE} = 0.583\ \mu V$) and J4 ($V_{SNE} = 0.6204\ \mu V$) clearly show a transverse SNE response as shown in Supplementary Figure S14 (b) and (d). The offset thermal response in the measurements is expected to arise due to Seebeck effect in Si, which has a large Seebeck coefficient.

Both SNE and PNE responses are much larger than the PNE response expected from $Ni_{80}Fe_{20}$ only, which again suggests interlayer spin-phonon coupling leading to enhanced magneto-thermal transport behavior. The observed SNE response cannot arise due to non-local current leakage since J4 is farther than J3 (where PNE response is observed). In addition, non-local current leakage will give rise to anisotropic thermopower response, which has a symmetry of $\sin^2 \theta$ instead of $\sin 2\theta$. Intriguingly, SNE response is only observed at junctions J2 and J4 while not at J3. This is attributed to local variation in strain gradient due to inhomogeneous thermal expansion. As a consequence, differences in interlayer coupling will arise, which will give rise to the observed behavior. The measured phononic SNE response is inconclusive since not all location shows the SNE behavior. However, this measurement conclusively shows the existence of interlayer spin-phonon coupling.